\documentclass[aps,prl,twocolumn,superscriptaddress,showpacs]{revtex4}
\usepackage{graphicx}

\def\nLUMI{78\fbinv}
\def\nBB{85.0\times10^{6}}
\def\nBBfull{(85.0\pm0.5)\times10^{6}}

\def\LcontCut{0.65}         

\def\EffKaon{86\%}
\def\EffPion{96\%}

\def\NumKPPMG{450.1\pm24.2\pm6.1}
\def\NumKSPPG{145.0\pm13.7\pm3.6}
\def\NumKPPZG{129.1\pm14.7\pm5.5}
\def\NumKSPZG{23.8\pm6.4\pm1.0}
\def\NumKZG{473.9\pm25.0\pm6.2}
\def\NumKPG{274.1\pm20.1\pm6.6}

\def\NDEBtoKPPMG{453.2\pm27.4}
\def\NDEBtoKSPPG{147.8\pm15.4}
\def\NDEBtoKPPZG{120.0\pm14.8}
\def\NDEBtoKSPZG{23.2\pm6.2}
\def\NDEBtoKZG{476.4\pm28.1}
\def\NDEBtoKPG{267.8\pm21.4}

\def\NKpiBtoKZG{480.6\pm25.7}
\def\NKpiBtoKPG{270.9\pm19.9}
\def\NnonresBtoKZG{5.9\pm2.3}
\def\NnonresBtoKPG{6.4\pm1.9}
\def\FracnonresZ{1.2\pm0.5} 
\def\SysnonresZ{1.3}        
\def\FracnonresP{2.4\pm0.4} 
\def\SysnonresP{2.4}        

\def\EffKPPMG{12.83\pm0.54}
\def\EffKSPPG{3.96\pm0.25}
\def\EffKPPZG{3.63\pm0.19}
\def\EffKSPZG{1.09\pm0.07}
\def\EffKZG{13.92\pm0.58}
\def\EffKPG{7.59\pm0.39}

\def\sysFracKPPM{4.2\%}
\def\sysFracKSPP{6.3\%}
\def\sysFracKPPZ{5.3\%}
\def\sysFracKSPZ{6.8\%}

\def\BrBtoKZG{4.01 \pm 0.21 \pm 0.17}
\def\BrBtoKPG{4.25 \pm 0.31 \pm 0.24}
\def\BrBtoKPPMG{4.13 \pm 0.22 \pm 0.18}
\def\BrBtoKSPPG{4.31 \pm 0.41 \pm 0.29}
\def\BrBtoKPPZG{4.19 \pm 0.48 \pm 0.28}
\def\BrBtoKSPZG{2.57 \pm 0.69 \pm 0.20}
\def\ResultDeltazp{+0.012 \pm 0.044 \pm 0.026} 
\def\ResultDeltazpFull{+0.012 \pm 0.044 {\rm(stat)}\pm 0.026{\rm(syst)}}

\def\NnegBtoKZG{218.5 \pm 16.8 \pm 3.0}
\def\NposBtoKZG{231.6 \pm 17.4 \pm 3.0}
\def\NnegBtoKSPPG{79.2 \pm 10.0 \pm 1.9}
\def\NposBtoKSPPG{65.8 \pm 9.4 \pm 1.9}
\def\NnegBtoKPPZG{68.7 \pm 11.6 \pm 2.2}
\def\NposBtoKPPZG{80.1 \pm 12.4 \pm 2.5}
\def\AcpBtoKZG{-0.030 \pm 0.055 \pm 0.014}
\def\AcpBtoKSPPG{+0.094 \pm 0.094 \pm 0.021}
\def\AcpBtoKPPZG{-0.078 \pm 0.113 \pm 0.028}
\def\AcpBtoKPG{+0.007 \pm 0.074 \pm 0.017}
\def\AcpBtoKG{-0.015 \pm 0.044 \pm 0.012}
\def\AcpBtoKGfull{-0.015 \pm 0.044\mathrm{(stat)} \pm 0.012\mathrm{(syst)}}

\def\ResoMbc{2.73\pm0.04}
\def\ResoMbcPZ{3.35\pm0.10}

\def\sysNBB{0.6\%}
\def\sysPhoton{2.2\%}
\def\sysDblTracking{2.0\%}
\def\sysTracking{1.0\%}
\def\sysKidKPPM{0.3\%}
\def\sysKidKPPZ{0.3\%}
\def\sysPiid{0.3\%}
\def\sysKS{4.5\%}
\def\sysPZ{2.7\%}
\def\sysMKpi{1.8\%}
\def\sysLR{1.8\%}
\def\sysMCKPPM{0.7\%}
\def\sysMCKSPP{1.2\%}
\def\sysMCKPPZ{1.5\%}
\def\sysMCKSPZ{2.4\%}
\def\BAR#1{\overline{#1}{}}

\def\qqbar{q\BAR{q}}

\def\Bbar{\BAR{B}}

\def\Kbar{\BAR{K}}

\def\piP{\pi^+}
\def\piM{\pi^-}
\def\piZ{\pi^0}

\def\KP{K^+}
\def\KM{K^-}

\def\KS{K^0_S}

\def\KstarP{K^{*+}}

\def\KstarZ{K^{*0}}

\def\Dzero{D^0}
\def\Dplus{D^+}

\def\Bzero{B^0}
\def\Bplus{B^+}

\def\KPPM{\KP\piM}
\def\KSPZ{\KS\piZ}
\def\KSPP{\KS\piP}
\def\KPPZ{\KP\piZ}

\def\epem{e^+e^-}

\def\piZeta{\piZ/\eta}  

\def\BtoKstargamma{B\to K^*\gamma}
\def\BtoKstarG{B\to K^*\gamma}

\def\BtoKstarZG{B^0\to K^{*0}\gamma}
\def\BtoKstarPG{B^+\to K^{*+}\gamma}

\def\BtoKG{\BtoKstarG}
\def\BtoKZG{\BtoKstarZG}
\def\BtoKPG{\BtoKstarPG}

\def\BtoDpi{B\to D\piM}
\def\BtoDZpi{B^-\to\Dzero\piM}
\def\BtoDPpi{\Bbar^0\to D^+\piM}
\def\DZtoKpi{\Dzero\to\KM\piP}
\def\DPtoKpipi{D^+\to\KM\piP\piP}

\def\Br{{\cal B}}
\def\Acp{A_{CP}}
\def\Deltazp{\Delta_{0+}}

\def\vecpp{\vec{p}^{\,}{}'}

\def\vecpB{\vec{p}^{\,*}_B}
\def\vecpgamma{\vec{p}^{\,*}_\gamma}
\def\vecpKstar{\vec{p}^{\,*}_{K^*}}

\def\thetahel{\theta_{\rm hel}}
\def\coshel{\cos\theta_{\rm hel}}
\def\sinsqhel{1-\cos^2\theta_{\rm hel}}
\def\cossqhel{\cos^2\theta_{\rm hel}}

\def\Mbc{M_{\rm bc}}

\def\MKpi{M(K\pi)}

\def\Ebeam{E^{\,*}_{\rm beam}{}}
\def\Egamma{E^{\,*}_\gamma}
\def\EKstar{E^{\,*}_{K^*}}
\def\EB{E_B^{\,*}{}}

\def\DeltaE{\Delta{E}}

\def\Lcont{{\cal L}_{\rm cont}}

\def\mm{\mbox{~mm}}
\def\cm{\mbox{~cm}}

\def\fbinv{\mbox{~fb}^{-1}}

\def\MeV{\mbox{~MeV}}
\def\MeVc{\MeV\!/c}
\def\MeVcc{\MeVc^2}
\def\GeV{\mbox{~GeV}}
\def\GeVc{\GeV\!/c}
\def\GeVcc{\GeVc^2}

\def\rad{\mbox{~rad}}

\def\Journal#1#2#3#4{{#1} {\bf #2}, #3 (#4)}
\def\NIMA{Nucl. Instrum. Meth. A}
\def\NPB{Nucl. Phys. B}
\def\PLB{Phys. Lett. B}
\def\PRL{Phys. Rev. Lett.}
\def\PRD{Phys. Rev. D}

\def\EPJC{Eur. Phys. J. C}
\def\etal{{\it et al.}}

\newcommand{\myfigure}[3]{%
  \begin{figure}[H]
  \begin{center}
    \includegraphics[width=#1\textwidth,keepaspectratio]{#2_color.eps}
    \caption{#3}\label{fig:#2}
  \end{center}\end{figure}}

\newenvironment{mytable}[3]{%
  \begin{table}[H]
  \begin{center}
    \caption{#3}\label{tbl:#2}
    \begin{tabular}{#1}}{
    \end{tabular}
  \end{center}\end{table}}

\newenvironment{mytable*}[3]{%
  \begin{table*}[H]
  \begin{center}
    \caption{#3}\label{tbl:#2}
    \begin{tabular}{#1}}{
    \end{tabular}
  \end{center}\end{table*}}

\begin{document}

\hbox to \textwidth{%
                 \resizebox{!}{3cm}{
                 \includegraphics{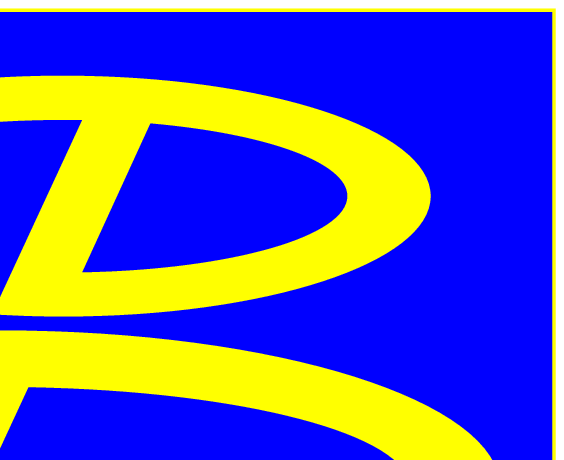}}
                 \hss{\vbox to 3cm{%
                 \hbox{Belle-Preprint-2004-5}
                 \hbox{KEK-Preprint 2003-132}\vss
}}}


\title{ \quad\\[1cm] \Large
Measurement of the $\BtoKG$ Branching Fractions and Asymmetries}


\affiliation{Budker Institute of Nuclear Physics, Novosibirsk}
\affiliation{Chiba University, Chiba}
\affiliation{University of Cincinnati, Cincinnati, Ohio 45221}
\affiliation{University of Frankfurt, Frankfurt}
\affiliation{University of Hawaii, Honolulu, Hawaii 96822}
\affiliation{High Energy Accelerator Research Organization (KEK), Tsukuba}
\affiliation{Hiroshima Institute of Technology, Hiroshima}
\affiliation{Institute of High Energy Physics, Chinese Academy of Sciences, Beijing}
\affiliation{Institute of High Energy Physics, Vienna}
\affiliation{Institute for Theoretical and Experimental Physics, Moscow}
\affiliation{J. Stefan Institute, Ljubljana}
\affiliation{Kanagawa University, Yokohama}
\affiliation{Korea University, Seoul}
\affiliation{Kyungpook National University, Taegu}
\affiliation{Swiss Federal Institute of Technology of Lausanne, EPFL, Lausanne}
\affiliation{University of Ljubljana, Ljubljana}
\affiliation{University of Maribor, Maribor}
\affiliation{University of Melbourne, Victoria}
\affiliation{Nagoya University, Nagoya}
\affiliation{Nara Women's University, Nara}
\affiliation{National United University, Miao Li}
\affiliation{Department of Physics, National Taiwan University, Taipei}
\affiliation{H. Niewodniczanski Institute of Nuclear Physics, Krakow}
\affiliation{Nihon Dental College, Niigata}
\affiliation{Niigata University, Niigata}
\affiliation{Osaka City University, Osaka}
\affiliation{Osaka University, Osaka}
\affiliation{Panjab University, Chandigarh}
\affiliation{Peking University, Beijing}
\affiliation{Princeton University, Princeton, New Jersey 08545}
\affiliation{RIKEN BNL Research Center, Upton, New York 11973}
\affiliation{University of Science and Technology of China, Hefei}
\affiliation{Seoul National University, Seoul}
\affiliation{Sungkyunkwan University, Suwon}
\affiliation{University of Sydney, Sydney NSW}
\affiliation{Tata Institute of Fundamental Research, Bombay}
\affiliation{Toho University, Funabashi}
\affiliation{Tohoku Gakuin University, Tagajo}
\affiliation{Tohoku University, Sendai}
\affiliation{Department of Physics, University of Tokyo, Tokyo}
\affiliation{Tokyo Institute of Technology, Tokyo}
\affiliation{Tokyo Metropolitan University, Tokyo}
\affiliation{Tokyo University of Agriculture and Technology, Tokyo}
\affiliation{Toyama National College of Maritime Technology, Toyama}
\affiliation{University of Tsukuba, Tsukuba}
\affiliation{Utkal University, Bhubaneswer}
\affiliation{Virginia Polytechnic Institute and State University, Blacksburg, Virginia 24061}
\affiliation{Yokkaichi University, Yokkaichi}
\affiliation{Yonsei University, Seoul}
  \author{M.~Nakao}\affiliation{High Energy Accelerator Research Organization (KEK), Tsukuba} 
  \author{K.~Abe}\affiliation{High Energy Accelerator Research Organization (KEK), Tsukuba} 
  \author{K.~Abe}\affiliation{Tohoku Gakuin University, Tagajo} 
  \author{T.~Abe}\affiliation{High Energy Accelerator Research Organization (KEK), Tsukuba} 
  \author{I.~Adachi}\affiliation{High Energy Accelerator Research Organization (KEK), Tsukuba} 
  \author{H.~Aihara}\affiliation{Department of Physics, University of Tokyo, Tokyo} 
  \author{M.~Akatsu}\affiliation{Nagoya University, Nagoya} 
  \author{Y.~Asano}\affiliation{University of Tsukuba, Tsukuba} 
  \author{T.~Aso}\affiliation{Toyama National College of Maritime Technology, Toyama} 
  \author{V.~Aulchenko}\affiliation{Budker Institute of Nuclear Physics, Novosibirsk} 
% ...
\collaboration{The Belle Collaboration}

\begin{abstract} 

We report measurements of the radiative decay $\BtoKG$.  The analysis is
based on a data sample containing $\nBB$ $B$ meson pairs collected by
the Belle detector at the KEKB storage ring.  We measure branching
fractions of $\Br(\BtoKZG)=(\BrBtoKZG)\times10^{-5}$ and
$\Br(\BtoKPG)=(\BrBtoKPG)\times10^{-5}$, where the first and second
errors are statistical and systematic, respectively.  The isospin
asymmetry between $\Bzero$ and $\Bplus$ decay widths is measured to be
$\Deltazp = \ResultDeltazp$.  We search for a partial rate asymmetry
between $CP$ conjugate modes, and find $\Acp(\BtoKG)=\AcpBtoKG$.
\end{abstract}
%
\pacs{13.40.Hq, 14.40.Nd}  

\maketitle


{\renewcommand{\thefootnote}{\fnsymbol{footnote}}


\normalsize
\setcounter{footnote}{0}
\newpage
\normalsize


                    \section{Introduction}

One decade after the first observation of exclusive $\BtoKG$ decays by
CLEO in 1993 \cite{bib:cleo-1993}, this process continues to be a
subject of considerable interest.  The size of the decay rate itself
provides only a mild constraint on extensions to the Standard Model
(SM), because SM predictions for exclusive rates suffer from large
(${\sim}30\%$) and model dependent form factor uncertainties
\cite{bib:kstargam-theories}.  Of more interest are asymmetries, where
theoretical uncertainties largely cancel.  The isospin asymmetry between
the charged and neutral $\BtoKG$ decay widths is predicted to be $+5$ to
$10\%$ in the SM, while in some SM extensions it may have an opposite
sign \cite{bib:kagan-neubert-2001}.  A measurement of the partial rate
asymmetry between $CP$ conjugate modes is another interesting subject;
here the SM expectation is much less than $1\%$ and any large asymmetry
would be an indication of non-SM effects.  In this report, we present
new measurements of the $\BtoKstargamma$ branching fractions, and
isospin and charge asymmetries.

                    \section{Dataset and Apparatus}

The data sample used in this analysis contains $\nBBfull$ $B$ meson
pairs, corresponding to an integrated luminosity of $\nLUMI$, collected
at the $\Upsilon(4S)$ resonance by the Belle detector at the KEKB
storage ring.  KEKB is a double-ring asymmetric-energy $\epem$ storage
ring ($3.5\GeV$ on $8 \GeV$) \cite{bib:kekb}.  We also use an
off-resonance data sample of $8.3\fbinv$ collected at a center-of-mass
(CM) energy that is $60\MeV$ below the $\Upsilon(4S)$ resonance.

The Belle detector is a large-solid-angle magnetic spectrometer that
consists of a three-layer silicon vertex detector (SVD), a 50-layer
central drift chamber (CDC), an array of aerogel threshold Cherenkov
counters (ACC), a barrel-like arrangement of time-of-flight
scintillation counters (TOF), and an electromagnetic calorimeter (ECL)
comprised of CsI(Tl) crystals located inside a super-conducting
solenoid coil that provides a 1.5~T magnetic field.  An iron flux-return
located outside of the coil is instrumented to detect $K_L^0$ mesons and
to identify muons (KLM).  The detector is described in detail
elsewhere~\cite{bib:belle-nim}.

                    \section{Signal Reconstruction}

The analysis is performed by reconstructing $B$ meson candidates that
include a high energy primary photon and a $K^*$ resonance reconstructed
in one of four final states: $\KPPM$, $\KSPZ$, $\KSPP$, and $\KPPZ$.
Here and throughout this report, $K^*$ denotes the $K^*(892)$, and the
inclusion of charge conjugate modes is implied unless otherwise
stated.

Photon ($\gamma$) candidates are reconstructed from isolated clusters in
the ECL that have no corresponding charged track, and a shower shape that
is consistent with that of a photon.  The photon energy is calculated
from the sum of energies in crystals with more than 0.5 MeV energy
deposited around the central cell.  Photons in the energy range
$1.8\GeV<\Egamma<3.4\GeV$ in the barrel region of the ECL
($33^\circ<\theta_\gamma<128^\circ$) are selected as primary photon
candidates from $B$ decay; here $\Egamma$ is the photon energy in the CM
frame and $\theta_\gamma$ is the polar angle in the laboratory frame.
(We use variables calculated both in the CM frame and laboratory frame:
variables defined in the CM frame are labeled with an asterisk.)  In
order to reduce the backgrounds from $\piZ$ and $\eta$ mesons from
continuum light quark-pair production ($\epem\to\qqbar$, $q=u,d,s,c$),
we impose two additional requirements on the primary photon.  One is the
explicit removal of $\piZ$ ($\eta$) candidates by requiring the
invariant mass of the primary photon and any other photon with an energy
greater than $30\MeV$ ($200\MeV$) to be outside of a window of
$\pm18\MeVcc$ ($\pm32\MeVcc$) around the nominal $\piZ$ ($\eta$) mass.
These correspond to $\pm3\sigma$ windows, where $\sigma$ is the mass
resolution.  This set of criteria is referred to as the $\piZ/\eta$
veto.  The other requirement is the removal of the clusters that are not
fully consistent with an isolated electromagnetic shower.  We require
the ratio of the energy deposition in $3\times3$ cells to that in
$5\times5$ cells around the maximum energy ECL cell of the cluster
($E_{9/25}$) to be greater than 0.95, which retains 95\% of the signal
photons.

Charged pions ($\piP$) and kaons ($\KP$) are reconstructed as tracks in
the CDC and SVD.  The tracks are required to originate from the
interaction region by requiring that they have radial impact parameters
relative to the run-averaged measured interaction point of less than
$1.5\cm$.  We determine the pion ($L_{\pi}$) and kaon ($L_K$)
likelihoods from the ACC response, the specific ionization ($dE/dx$)
measurement in the CDC and the TOF flight-time measurement for each
track, and form a likelihood ratio $L_{K/\pi}=L_K/(L_\pi + L_K)$ to
separate pions and kaons.  We require $L_{K/\pi} > 0.6$ for kaons, which
gives an efficiency of $\EffKaon$ for kaons, and $L_{K/\pi} < 0.9$ for
pions, which gives an efficiency of $\EffPion$ for pions.  In addition,
we remove kaon and pion candidates if they are consistent with being
electrons based on the ECL, $dE/dx$, and ACC information.

Neutral pions ($\piZ$) are formed from two photons with invariant masses
within $\pm 16\MeV$ ($3\sigma$) of the $\piZ$ mass; the photon momenta
are then recalculated with a $\piZ$ mass constraint.  The $\piZ$ mass
resolution is better than that for the $\piZeta$ veto where the photon
energies are highly asymmetric.  We require each photon energy to be
greater than 50 MeV, and the cosine of the angle between the two photons
($\cos\theta_{\gamma\gamma})$ to be greater than 0.5.  This angle
requirement is almost equivalent to selecting $\piZ$s with momentum above
$0.5\GeVc$; it retains about 90\% of the signal $\piZ$s while rejecting
43\% of the $\piZ$ candidates in the background.

Neutral kaons ($\KS$) are reconstructed from two oppositely charged
pions that have invariant masses within $\pm10\MeV$ $(3\sigma)$ of the
$\KS$ mass; the pion momenta are then recalculated with a $\KS$ vertex
constraint.  We impose additional criteria based on the radial impact
parameters of the pions ($\delta r$), the distance between the closest
approaches of the pions along the beam direction ($\delta z$), the
distance of the vertex from the interaction point ($l$), and the
azimuthal angle difference between the vertex direction and the $\KS$
momentum direction ($\delta\phi$).  These variables are combined as
follows: for $p(\KS)<0.5\GeVc$, $\delta z<8\mm$, $\delta r>0.5\mm$, and
$\delta\phi<0.3\rad$ are required; for $0.5\GeVc<p(\KS)<1.5\GeVc$, $\delta
z<18\mm$, $\delta r>0.3\mm$, $\delta\phi<0.1\rad$, and $l>0.8\mm$ are
required; and for $p(\KS)>1.5\GeVc$, $\delta z<24\mm$, $\delta
r>0.2\mm$, $\delta\phi<0.03\rad$, and $l>2.2\mm$ are required.  This set
of criteria retains about 80\% of the signal $\KS$.

We form a $B$ candidate from a primary photon candidate and a $K^*$
candidate, which is a $K\pi$ system with an invariant mass $\MKpi$
within $\pm75\MeVcc$ of the $K^*$ mass.  In order to separate the $B$
candidate from backgrounds, we form two kinematic variables: the
beam-energy constrained mass $\Mbc=\sqrt{(\Ebeam/c^2)^2 - |\vecpB/c|^2}$
and the energy difference $\DeltaE=\EB - \Ebeam$, where $\Ebeam$ is the
beam energy, and $\EB$ and $\vecpB$ are the energy and momentum,
respectively, of the $B$ candidate in the CM frame.  The energy $\EB$ is
calculated as $\EB=\Egamma+\EKstar$; the momentum $\vecpB$ is calculated
without using the absolute value of the photon momentum according to
\begin{equation}
\vecpB=\vecpKstar + {\vecpgamma\over|\vecpgamma|} \times (\Ebeam - \EKstar),
\end{equation}
since the $K^*$ momentum and the beam energy are determined with
substantially better precision than that of the primary photon.

We use $\Mbc$ as the primary distribution to extract the signal yield.
For modes without a $\piZ$, we use a Gaussian function with a width of
$(\ResoMbc)\MeVcc$ to model the signal; for modes with a $\piZ$, we use
an empirical formula to reproduce the asymmetric ECL energy response
(known as the Crystal Ball line shape \cite{bib:crystal-ball}), whose
effective width is $(\ResoMbcPZ)\MeVcc$.  The peak positions and widths
are primarily determined using Monte Carlo (MC) samples, and corrected
for the measured differences in the beam-energy and its spread between
data and MC using a $B^-\to D^0\pi^-$ sample.  The $\DeltaE$ signal
distribution also has a large tail on the negative $\DeltaE$ side due to
the asymmetric ECL energy response.  For the modes without a $\piZ$, we
use a Crystal Ball line shape; for the modes with a $\piZ$, we convolve
an additional Gaussian resolution function to describe a broader width
and add a broad Gaussian component for the small tail in the positive
$\DeltaE$ side.  These shapes are determined using MC samples.  We
select candidates with $-200\MeV<\DeltaE<100\MeV$ to accommodate the
asymmetric $\DeltaE$ signal shape.  We define a $\DeltaE$ sideband as
$100\MeV<\DeltaE<400\MeV$, where no signal is expected, to study the
$\Mbc$ distribution of the background.
There is no background that makes a peak in this $\DeltaE$ sideband.
We require $\Mbc>5.270\GeVcc$ when the $\DeltaE$ distribution is
examined.  We define $5.227\GeVcc<\Mbc<5.263\GeVcc$ as an $\Mbc$
sideband that is used to study the $\DeltaE$ distribution of the
background.

                    \section{Background Suppression}

The main background source is continuum $\qqbar$ production including
the initial state radiation process $\epem\to \qqbar\gamma$.  We reduce
this background by exploiting the topological event shape differences:
$B$ meson pairs decay almost at rest in the CM frame and thus the final
state particles are distributed nearly isotropically; $\qqbar$ pairs are
produced back-to-back with multi-$\mbox{GeV}/c$ momenta in both
hemispheres and, thus, tend to be more two-jet like.

We define a Fisher descriminant ($F$) \cite{bib:fisher} from modified
Fox-Wolfram moments \cite{bib:fox-wolfram},
\begin{equation}
F=\alpha_2 R_2^{so} + \alpha_4 R_4^{so} + \sum_{l=1}^4 \beta_l R_l^{oo},
\end{equation}
where $\alpha_l$, $\beta_l$ are coefficients that are selected to
provide the maximum discrimination between the signal and the continuum
background.  The modified Fox-Wolfram moments are defined as
\begin{equation}
\begin{array}{l}
\displaystyle
R_l^{so}={\sum_{i,\gamma} |\vecpp_i||\vecpp_\gamma| P_l(\cos\theta'_{i\gamma})
    \over \sum_{i,\gamma} |\vecpp_i||\vecpp_\gamma|},\\[18pt]  
\displaystyle
R_l^{oo}={\sum_{i,j} |\vecpp_i||\vecpp_j| P_l(\cos\theta'_{ij})
    \over \sum_{i,j} |\vecpp_i||\vecpp_j|},\\            
\end{array}
\end{equation}
where the indices $i$, $j$ indicate the charged tracks (with a $\pi^+$
mass hypothesis) and photons that are not used to form the $B$
candidate, and the index $\gamma$ corresponds to the primary photon.
The variables $\theta'_{ij}$ and $\theta'_{i\gamma}$ are the opening
angles between two momentum vectors, and $P_l$ is the $l$-th Legendre
polynomial function.  The momenta ($\vecpp$) and angles ($\theta'$) are
calculated in the candidate $B$ rest frame (denoted as primed
variables), since the selection efficiency with this variable has a
smaller correlation with $\Mbc$ than that calculated in the CM frame.

As an additional discriminant, we use the cosine of the CM polar angle
of the candidate $B$ flight direction, $\cos\theta^*_B$.  The
$\cos\theta^*_B$ distribution is $1-\cos^2\theta^*_B$ for $B$ production
from $e^+e^-\to\Upsilon(4S)$, and is found to be flat for the continuum
background.

We combine these two discriminants into a likelihood ratio,
\begin{equation}
\begin{array}{c}
\displaystyle
\Lcont={L_S \over L_S + L_B},
\\[18pt]
\displaystyle
L_S=P^F_S \times P^{\rm cosB}_S\mbox{,~~~}
L_B=P^F_B \times P^{\rm cosB}_B,
\end{array}
\end{equation}
where $P^F$ and $P^{\rm cosB}$ are the probability density functions
(PDF) for the Fisher and the $B$ flight direction, and the indices $S$ and
$B$ denote the signal and background.  For $P^{\rm cosB}_S$ and $P^{\rm
cosB}_B$, we use ${3\over2}(1-\cos^2\theta^*_B)$ and ${1\over2}$,
respectively.  For $P^F_S$ and $P^F_B$, we model the shape by fitting
the signal and continuum background MC distributions with
asymmetric Gaussian functions for each of the four $\BtoKG$ channels.

The value of $\Lcont$ ranges between 0 and 1.  We optimize the minimum
$\Lcont$ requirement to provide the largest value of
$N_S/\sqrt{N_S+N_B}$, where $N_S$ and $N_B$ are the expected signal and
background yields for $\nBB$ $B$ meson pairs assuming previously
measured $B\to K^*\gamma$ branching fraction values
\cite{bib:belle-old,bib:cleo-2000,bib:babar-2002}.  Although there is a
slight mode dependence in the optimal value, we apply the same
requirement, $\Lcont>\LcontCut$, which is close to the optimal point for
each mode.  This requirement retains 73\% of signal events while
rejecting 90\% of continuum background events.

The remaining continuum background is distinguished by fits to the
$\Mbc$ distribution.  The continuum background is modeled with a
threshold function (known as the ARGUS function
\cite{bib:argus-function}),
\begin{equation}
\begin{array}{lcl}
\displaystyle
f_{\rm cont}(\Mbc)
   &=& \displaystyle N \times \Mbc
       \times \sqrt{1 - \left(\Mbc\over \Ebeam\right)^2} \\[18pt]
   & & \displaystyle \times \exp\left\{\alpha\left[1
                      - \left(\Mbc\over \Ebeam\right)^2\right]\right\},
\end{array}
\end{equation}
where $N$ is a normalization factor and $\alpha$ is an empirical shape
parameter.  We determine the shape parameter from the $\DeltaE$ sideband
data, since there is no significant difference between the background
shapes in the $\DeltaE$ sideband and the signal region for off-resonance
data and MC events.  The results are consistent for data and MC, and for
the different $\BtoKG$ channels.  We use the same background shape
parameter $\alpha$ for all the $\BtoKG$ channels.  For $\DeltaE$, the
background shape is parameterized as a linear function and determined
from $\Mbc$ sideband data.

Major background contributions from $B$ decays are from cross-feeds
between charged and neutral $B\to K^*\gamma$ decays, $B\to (K^*\pi +
K\rho)\gamma$ \cite{bib:kxgam-prl}, $B\to K^*\eta$ \cite{bib:hfag} and
the unmeasured mode $B\to K^*\pi^0$ \cite{bib:pdg2003} for which we
assume half of the upper limit as the branching fraction with a 100\%
error.  These backgrounds peak in $\Mbc$ around the signal with a
slightly larger tail, and have a broad peak in $\DeltaE$ at negative
values.  We model these backgrounds and other $B$ decay backgrounds with
a smoothed histogram generated from a large MC sample.

                    \section{Signal extraction}

We extract signal yields in each of the four final states using a
one-dimensional binned likelihood fit.  The $\Mbc$ distributions are
modeled as a sum of three components: the signal, the continuum
background and the $B$ decay backgrounds that are described in the
previous sections.  Figure~\ref{fig:fig1_mbc} shows the result of the
fits; clear signals are seen in all four final states.  The size of the
$B$ decay background component, which is seen as a slight enhancement of
the background shape under the signal peak, is fixed in the fitting
procedure.  We vary the $B$ decay background components by the errors on
their branching fractions to evaluate the systematic error due to their
uncertainties.  We also vary the continuum background shape, $\Ebeam$,
the $B$ meson mass, and the $\Mbc$ resolution by their errors to
evaluate the systematic errors on the signal and background PDFs.  We
use a quadratic sum of the variations in the signal yield with these
tests as the systematic error on the signal yield.

A similar fitting procedure is performed for the $\DeltaE$ distributions
as shown in Fig.~\ref{fig:fig2_de}.  The $\DeltaE$ yields are obtained
by integrating the fit results from $-0.2\GeV$ to $0.1\GeV$ to allow a
comparison with the $\Mbc$ results.  The fit result for each mode is in
agreement with the $\Mbc$ result as given in Table~\ref{tbl:yield}.  In
this analysis, the $\DeltaE$ distribution is not as reliable as $\Mbc$
since we have to consider a wide $\DeltaE$ range where the background
contribution from $B$ decays is larger than that for $\Mbc$, and
$\DeltaE$ shapes for both signal and background may have large
uncertainties that are not fully evaluated due to lack of suitable
control samples.  Thus we base our signal yields on the $\Mbc$ fits.

The $\MKpi$ invariant mass spectrum, before applying the
$|\MKpi-M_{K^*}|<75\MeVcc$ requirement, gives discrimination of the
$K^*$ signal from resonances such as $K_2^*(1430)$ and $K^*(1410)$, or a
non-resonant $K\pi\gamma$ component.  
In order to examine the spectrum, we divide the data below
$M(K\pi)=2.0\GeVcc$ into $50\MeVcc$ wide bins and extract the signal
yield for each bin from a fit to the $\Mbc$ distribution, using the same
fitting procedure described above.
We veto the $D^0\to\KM\piP$ and $D^0\to\KS\piZ$ contributions from
$\Bbar^0\to D^0\pi^0$ and $\Bbar^0\to D^0\eta$ backgrounds by requiring
$|\MKpi-M_{\Dzero}|>20\MeVcc$ for the $\KstarZ\gamma$ modes.  Other $B$
decay backgrounds that may peak in $\Mbc$ are included as a background
component in each fit.  A $\chi^2$ fit is then performed to the $\MKpi$
spectrum using a sum of $K^*$, $K_2^*(1430)$ and $K^*(1410)$ resonances
and a non-resonant component without taking into account possible
interference effects.  The fit results are shown in
Fig.~\ref{fig:fig3_mkpi}.  For these resonances, we use relativistic
Breit-Wigner functions with the nominal masses and widths convolved with
a Gaussian resolution function.  In addition to the $K^*$ peak, we find
a significant $K_2^*(1430)$ peak, while the $K^*(1410)$ component is
consistent with zero.  The $K^*$ signal yields within the $\pm75\MeVcc$
window are consistent with the $\Mbc$ results as given in
Table~\ref{tbl:yield}.  In all cases, the background contributions from
$K^*(1410)$ and $K_2^*(1430)$ within the $K^*$ mass window are less than
one event including their errors.  The non-resonant contribution is
modeled with the inclusive $X_s$ mass spectrum of
Ref.~\cite{bib:kagan-neubert-1999}.  We find $\NnonresBtoKZG$ and
$\NnonresBtoKPG$ events under the peak of $K^{*0}$ and $K^{*+}$,
corresponding to $(\FracnonresZ)\%$ and $(\FracnonresP)\%$ of the signal
yields, respectively.  We include these yields into the systematic
errors instead of subtracting the contributions, since they could also
be due to a bias from other high mass resonances such as $K^*(1680)$
that hardly contribute to the $K^*$ mass peak.

The $K^*$ decay helicity angle, $\thetahel$, defined as the angle
between the kaon and the $B$ meson directions in the rest frame of the
$K\pi$ system, provides discrimination of the spin-1 signal from other
spin states.  The $\coshel$ distribution, shown in
Fig.~\ref{fig:fig4_coshel}, is obtained by dividing the data into bins
of 0.1 in $\coshel$ and fitting their $\Mbc$ distributions.  The $B$
decay backgrounds that peak in $\Mbc$ and follow a $\cossqhel$ structure
due to the pseudoscalar to pseudoscalar-vector nature of the decay are
included as a background component in each fit.  The $K^*\gamma$ signal
has a $\sinsqhel$ distribution.  In order to model a slight distortion
due to non-uniform tracking, particle identification and $\piZ$/$\KS$
reconstruction efficiencies, we use a fourth order polynomial function
that is constrained to zero at $\coshel=\pm1$.  We also add a flat
component modified with the same slight distortion, which turns out to
be consistent with zero.  The fit result is in agreement with the spin-1
signal.

We obtain a consistent set of signal yields from three different
distributions, $\Mbc$, $\DeltaE$, $\MKpi$ as summarized in
Table~\ref{tbl:yield}.  Both the $\MKpi$ and $\coshel$ distributions
suggest that the contributions from other resonances and non-resonant
decays can be neglected.  We conclude that the signal yields obtained
with the $\Mbc$ fit are essentially entirely due to $\BtoKG$ decays.

                    \section{Branching Fractions}

The reconstruction efficiencies are primarily obtained from signal MC
samples.  The selection criteria are divided into ten categories, and
the systematic error for each of them is evaluated with an independent
control sample.  The results are summarized in
Table~\ref{tbl:efficiency}.

The uncertainty in the photon detection efficiency is evaluated with a
sample of radiative Bhabha events.  For the tracking efficiency, we
quote an error from a comparison of the partially reconstructed
$D^{*+}\to\Dzero\piP$, $\Dzero\to\KS\piP\piM$ yield with the fully
reconstructed one.  For the charged kaon identification, we evaluate the
systematic error from a comparison between the efficiencies obtained
from kinematically selected $D^{*+}\to\Dzero\piP$, $\Dzero\to\KM\piP$
and $\phi\to\KP\KM$ decays in data and MC.  Similarly for the charged
pion identification, we compare the efficiency for the same $D^{*+}$
sample with that obtained from $\KS\to\piP\piM$ to evaluate the
systematic error.
The uncertainty in the $\KS$ reconstruction is obtained from a
comparison between $\Dplus\to\KS\piP$ with $\Dplus\to\KM\piP\piP$ in
$D^{*+}\to\Dplus\piZ$ decays.  The uncertainty in the $\piZ$
reconstruction efficiency is evaluated from a comparison of
$\eta\to\piZ\piZ\piZ$ to $\eta\to\gamma\gamma$ and $\eta\to\piP\piM\piZ$
in data and MC.  The efficiencies for the $\piZ/\eta$ veto and the
likelihood ratio requirement are evaluated together, using a $\BtoDpi$
control sample that includes the decay channels $\BtoDZpi$, $\DZtoKpi$
and $\BtoDPpi$, $\DPtoKpipi$.  We calculate $\Mbc$ as we do for
$\BtoKG$, {i.e.} without using the absolute value of the $\pi^-$
momentum.  For the $\piZ/\eta$ veto, we assume the $\pi^-$ is a massless
particle and scale the momentum by a factor of 1.1 to make the average
$\pi^-$ momentum equal that of the photons for $\BtoKG$.  We then
combine this massless $\pi^-$ with all photon candidates and reject the
event if the same $\piZ/\eta$ veto criteria are satisfied, and compare
the results between data and MC control samples.  The effect of the
$\MKpi$ requirement is evaluated by taking into account the $K^*$ form
factor and the detector resolution effects into the Breit-Wigner shape.

Using these reconstruction efficiencies, we obtain the branching
fractions for each of the four modes as summarized in
Table~\ref{tbl:branching}.  We add the signal yields and the
efficiencies for two modes of each of neutral and charged $B$ decays,
and obtain
\begin{equation}
\begin{array}{l@{=}l}
\displaystyle
\Br(B\to K^{*0}\gamma) & (\BrBtoKZG)\times 10^{-5} \\
\Br(B\to K^{*+}\gamma) & (\BrBtoKPG)\times 10^{-5}, \\
\end{array}
\end{equation}
where the first and the second errors are statistical and
systematic, respectively.  We assume an equal production rate for
$B^0\Bbar^0$ and $B^+B^-$ from the $\Upsilon(4S)$ resonance.

The isospin asymmetry,
\begin{equation}
\Deltazp = {
  (\tau_{B^+}/\tau_{B^0})\Br(\BtoKZG) - \Br(\BtoKPG) \over
  (\tau_{B^+}/\tau_{B^0})\Br(\BtoKZG) + \Br(\BtoKPG)},
\end{equation}
is then calculated from these results.  We use the world average value
of $\tau_{B^+}/\tau_{B^0} = 1.086\pm0.017$ \cite{bib:hfag}.  We assume
that the systematic error on the photon detection cancels.  The
systematic error on the $\Lcont$ and $\piZeta$ veto requirements is
estimated to be 0.013 in $\Deltazp$ from a comparison of the $\Bbar^0$
and $B^-$ subsets of the $\BtoDpi$ control sample.  We find the
systematic error due to the $\MKpi$ requirement is negligible.
Correlations between the systematic errors for the charged pion and kaon
tracking and particle identification, and the $\piZ$ and $\KS$
reconstruction efficiencies are taken into account.  Systematic errors
on the fitting procedures are assumed to be uncorrelated.

The result is
\begin{equation}
\Deltazp = \ResultDeltazpFull,
\end{equation}
which is consistent both with the SM prediction and no asymmetry.
Although the systematic errors in the branching fractions are almost as
large as the statistical errors, the systematic errors largely cancel in
$\Deltazp$.

If we allow the $B^+$ to $B^0$ production ratio ($f_+/f_0$) to
deviate from unity, the value of $\Deltazp$ shifts approximately by
${1\over2}(f_+/f_0-1)$.  The value of $f_+/f_0 = 1.044\pm0.050$ in
Ref.~\cite{bib:hfag} gives $\Deltazp =
+0.034\pm0.044\mbox{(stat)}\pm0.026\mbox{(syst)}\pm0.025(f_+/f_0)$.  The
conclusion above is therefore unchanged, although the result is shifted
closer to the SM prediction.

                    \section{Search for Partial Rate Asymmetry}

We define the partial rate asymmetry between $CP$ conjugate modes
(except for the $\KS\piZ\gamma$ mode) as
\begin{equation}
\begin{array}{lcl}
\Acp&=& \displaystyle
       {\Gamma(\Bbar \to \Kbar^{*}\gamma) - \Gamma(B \to K^*\gamma) \over
        \Gamma(\Bbar \to \Kbar^{*}\gamma) + \Gamma(B \to K^*\gamma) }\\[18pt]
    &=& \displaystyle
       {1\over 1-2w} \times
       {N(\Bbar \to \Kbar^{*}\gamma) - N(B \to K^*\gamma) \over
        N(\Bbar \to \Kbar^{*}\gamma) + N(B \to K^*\gamma) },
\end{array}
\end{equation}
where $N$ is the signal yield, $w$ is the wrong-tag fraction, $B$
indicates either $B^0$ or $B^+$, $K^*$ indicates $K^{*0}$($\to\KPPM$) or
$K^{*+}$($\to\KSPP,\KPPZ$), and $\Bbar$, $\Kbar^*$ are their conjugates,
respectively.

The wrong-tag fraction is negligible for $K^{*+}$, and is $0.9\%$
for $K^{*0}$ due to the double mis-identification of $\pi^+$ as $K^+$
and $K^-$ as $\pi^-$.  The wrong-tag fraction is obtained from the
signal MC; we neglect the small error on this fraction.

Possible detector and reconstruction biases are studied with an
inclusive $K^*$ sample.  We compare the yield of $K^*$ and $\Kbar^*$
and find no significant difference in
$A_{K^*}={1\over(1-2w)}{N(\Kbar^*)-N(K^*)\over N(\Kbar^*)+N(K^*)}$.  We
conclude there is no bias and assign systematic errors of 0.007 for
$\KstarZ\to\KP\piM$ and $\KstarP\to\KS\piP$, and 0.015 for
$\KstarP\to\KP\piZ$.  The systematic error on the $\Lcont$ and $\piZeta$
veto requirements is estimated to be 0.007 in $\Acp$ from a comparison
of the $B$ and $\Bbar$ subsets of the $\BtoDpi$ control sample.

We divide the data shown in each of Fig.~\ref{fig:fig1_mbc}(a--c) into
$CP$ conjugate modes.  We fit the $\Mbc$ distributions for the six modes
separately, using the same fitting procedure used in the branching
fraction measurement.  The fit results are shown in
Fig.~\ref{fig:fig5_mbcacp} and summarized in Table~\ref{tbl:acp}.  The
errors on the yield extraction are assumed to be uncorrelated.  By
assuming the partial rate asymmetry is equal for charged and neutral $B$
decays, we add the signal yields for $B$ and $\Bbar$ with a small
correction due to the wrong-tag fraction, and obtain
\begin{equation}
\Acp(\BtoKG)=\AcpBtoKGfull.
\end{equation}
Here the systematic error includes the errors on $A_{K^*}$, $\Lcont$ and
the yield extraction.

                    \section{Conclusions}

We have presented branching fraction, isospin asymmetry and $CP$ asymmetry
measurements for the radiative decay $\BtoKG$ using $\nBB$ $B$ meson
pairs.  The branching fraction results are consistent with previous
Belle \cite{bib:belle-old} results, and also with the CLEO
\cite{bib:cleo-2000} and BaBar \cite{bib:babar-2002} results, with the
errors improved by a factor of two.  The $K\pi$ mass spectra and the
decay helicity angle distribution in the $K^*$ mass region
($|\MKpi-M_{K^*}|<75\MeVcc$) are consistent with the dominance of $B\to
K^*\gamma$ without other contributions such as $K^*(1410)\gamma$ or
non-resonant $K\pi\gamma$ decays.  
We measure an isospin asymmetry which is consistent with zero; with the
current precision, our result agrees with both the SM prediction
and new physics scenarios which have the opposite sign.  For the partial
rate asymmetry between $CP$ conjugate modes, we obtain a result which is
also consistent with zero.  For both of these asymmetries, the
systematic errors are much smaller than the statistical errors, and
hence we can expect further improvements with larger data samples.

                    \section*{Acknowledgments}

We wish to thank the KEKB accelerator group for the excellent
operation of the KEKB accelerator.
We acknowledge support from the Ministry of Education,
Culture, Sports, Science, and Technology of Japan
and the Japan Society for the Promotion of Science;
the Australian Research Council
and the Australian Department of Education, Science and Training;
the National Science Foundation of China under contract No.~10175071;
the Department of Science and Technology of India;
the BK21 program of the Ministry of Education of Korea
and the CHEP SRC program of the Korea Science and Engineering
Foundation;
the Polish State Committee for Scientific Research
under contract No.~2P03B 01324;
the Ministry of Science and Technology of the Russian Federation;
the Ministry of Education, Science and Sport of the Republic of
Slovenia;
the National Science Council and the Ministry of Education of Taiwan;
and the U.S.\ Department of Energy.

\clearpage

\myfigure{0.5}{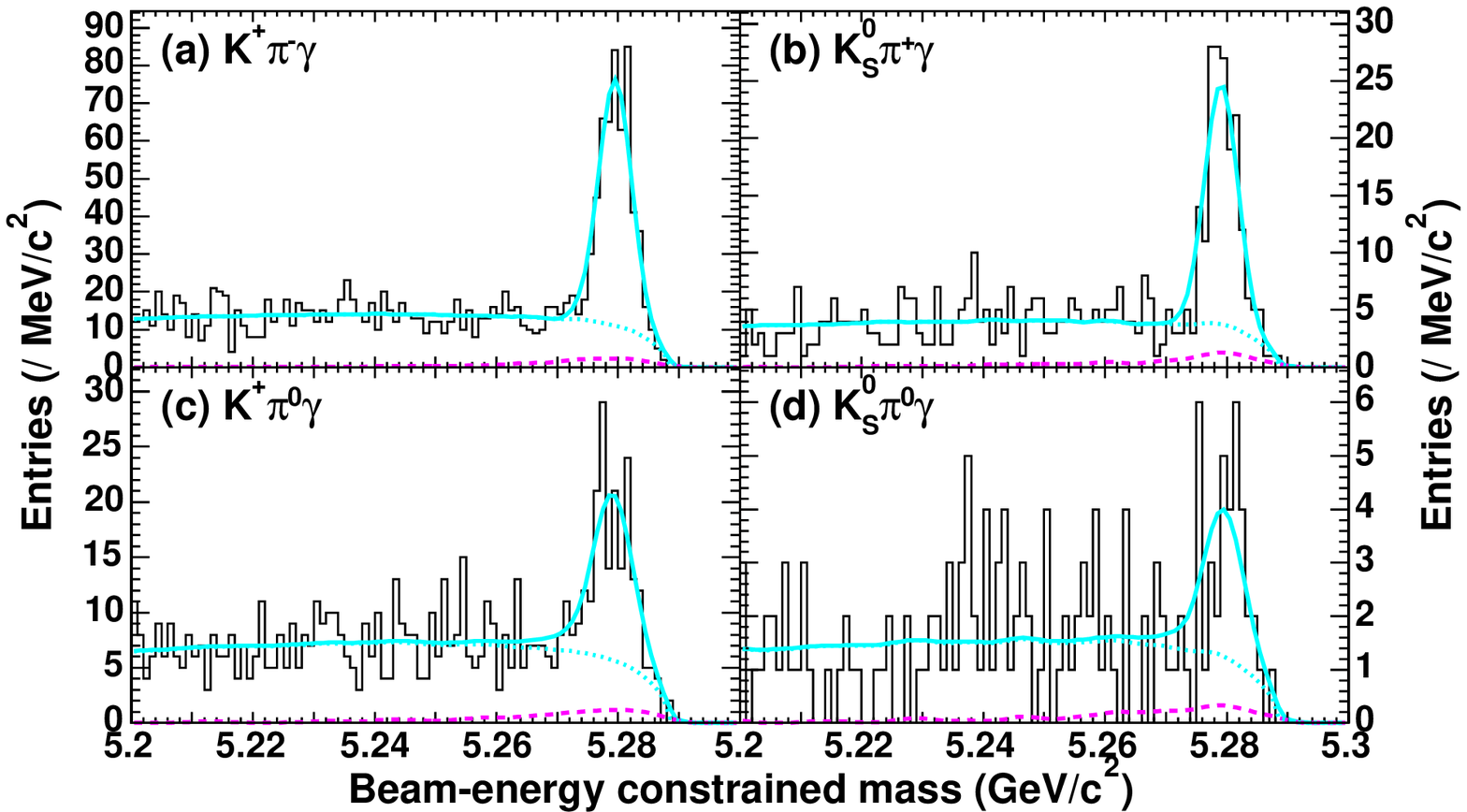}{%
  Fit results for the beam-energy constrained mass distribution for the
  (a) $\KPPM\gamma$, (b) $\KSPP\gamma$, (c) $\KPPZ\gamma$, and (d)
  $\KSPZ\gamma$ modes.  The sum of the signal and the background
  components are shown in the solid curves, while the dotted and dashed
  curves represent the total backgrounds and the $B$ decay backgrounds,
  respectively.}

\myfigure{0.5}{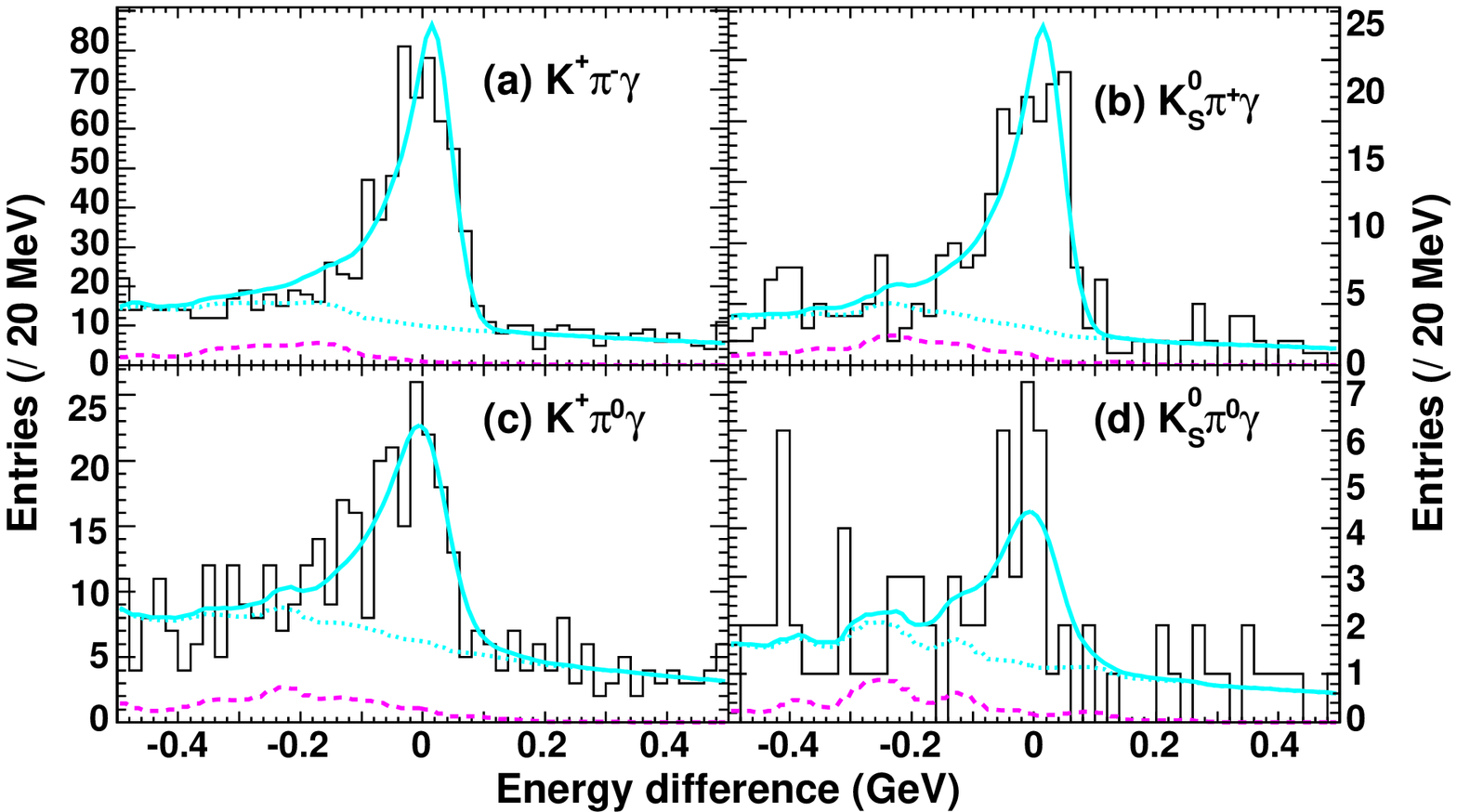}{%
  Fit results for the $\DeltaE$ distributions.  The sum of the signal
  and the background components are shown in the solid curves, while the
  dotted and dashed curves represent the total backgrounds and the $B$
  decay backgrounds, respectively.  }

\myfigure{0.5}{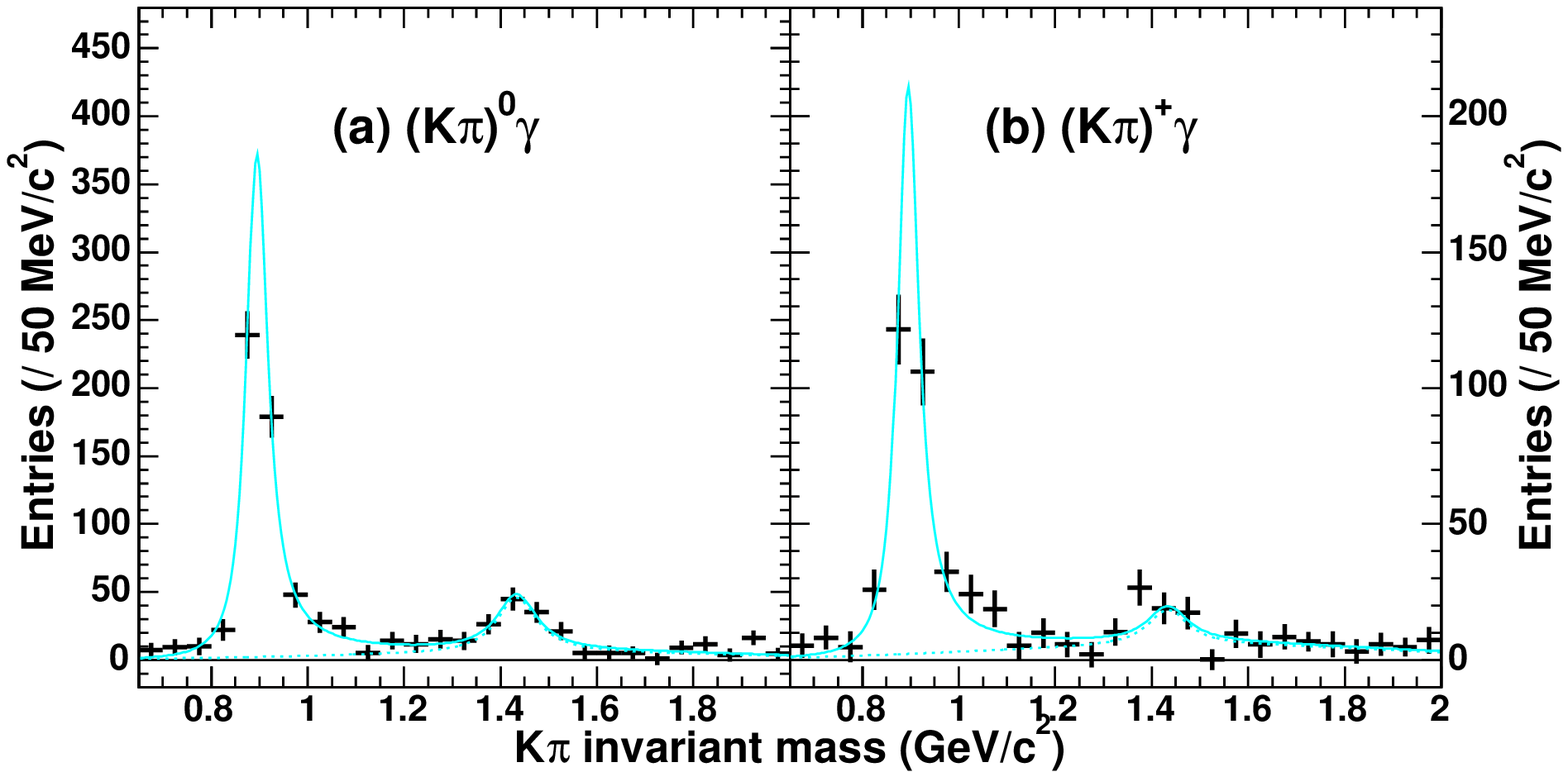}{%
  Fit results for the $K\pi$ invariant mass distributions for the (a)
  sum of $\KPPM\gamma$ and $\KSPZ\gamma$, and (b) sum of $\KSPP\gamma$
  and $\KPPZ\gamma$ channels.  The sum of the signal
  and the background components are shown in the solid curves, while the
  dotted curves represent the background components. }

\begin{mytable}{lccc}{yield}{Summary of the signal yields.}
\hline 
Mode & $\Mbc$ & $\DeltaE$ & $\MKpi$ \\
\hline 
$\KPPM\gamma$ & $\NumKPPMG$ & $\NDEBtoKPPMG$\\
$\KSPZ\gamma$ & $\NumKSPZG$ & $\NDEBtoKSPZG$\\
\hline 
$\KSPP\gamma$ & $\NumKSPPG$ & $\NDEBtoKSPPG$\\
$\KPPZ\gamma$ & $\NumKPPZG$ & $\NDEBtoKPPZG$\\
\hline 
$K^{*0}\gamma$   & $\NumKZG$ & $\NDEBtoKZG$ & $\NKpiBtoKZG$ \\
$K^{*+}\gamma$   & $\NumKPG$ & $\NDEBtoKPG$ & $\NKpiBtoKPG$ \\
\hline 
\end{mytable}

\myfigure{0.35}{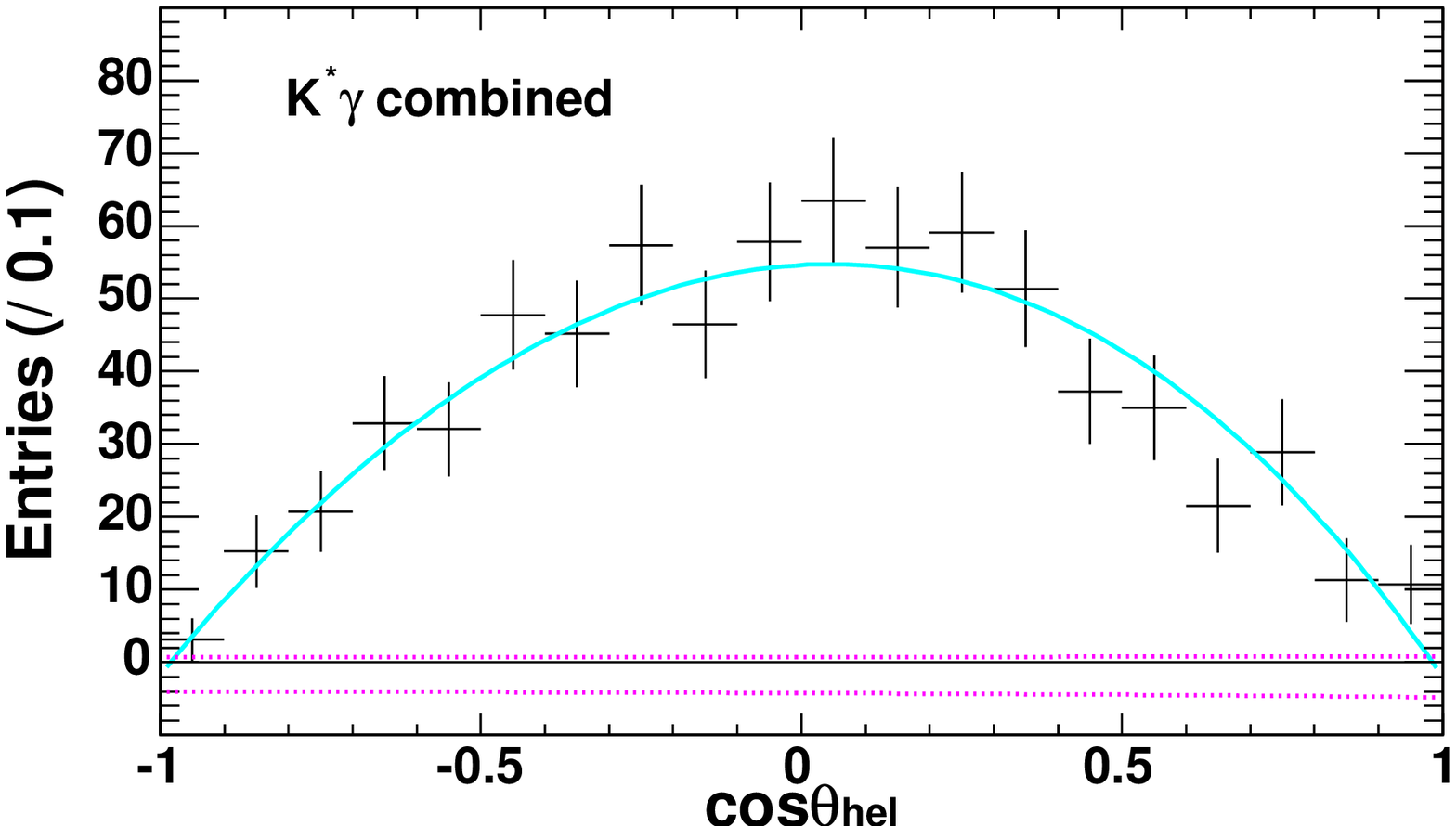}{%
  Fit results for the cosine of the helicity angle ($\coshel$)
  distribution for the sum of all four $\BtoKG$ channels.
  The solid line shows the sum of the signal and the flat
  component; the $\pm1\sigma$ bounds of the latter component
  are shown as the dotted lines. }

\begin{mytable*}{lcccc}{efficiency}{Reconstruction efficiencies and
 their systematic uncertainties. }
\hline 
{}     & $\KPPM\gamma$ & $\KSPP\gamma$ & $\KPPZ\gamma$ & $\KSPZ\gamma$ \\
\hline 
Reconstruction efficiency &
 $(\EffKPPMG)\%$ & $(\EffKSPPG)\%$ & $(\EffKPPZG)\%$ & $(\EffKSPZG)\%$ \\
\hline 
Fractional errors
        & $\sysFracKPPM$ & $\sysFracKSPP$ & $\sysFracKPPZ$ & $\sysFracKSPZ$ \\
~~~Number of $B$ meson pairs & $\sysNBB$ & $\sysNBB$ & $\sysNBB$ & $\sysNBB$ \\
~~~Photon selection
                 & $\sysPhoton$ & $\sysPhoton$ & $\sysPhoton$ & $\sysPhoton$ \\
~~~Tracking      & $\sysDblTracking$ & $\sysTracking$ & $\sysTracking$ & --- \\
~~~$K^+$ identification     & $\sysKidKPPM$ & --- & $\sysKidKPPZ$ & --- \\
~~~$\pi^-$ identification   & $\sysPiid$ & $\sysPiid$ & --- & --- \\
~~~$\KS$         & --- & $\sysKS$ & --- & $\sysKS$ \\
~~~$\pi^0$       & --- & --- & $\sysPZ$ & $\sysPZ$ \\
~~~$\Lcont+\pi^0/\eta$ veto     & $\sysLR$ & $\sysLR$ & $\sysLR$ & $\sysLR$ \\
~~~$\MKpi$       & $\sysMKpi$ & $\sysMKpi$ & $\sysMKpi$ & $\sysMKpi$ \\
~~~Non-resonant  & $\SysnonresZ\%$ & $\SysnonresP\%$ 
                 & $\SysnonresP\%$ & $\SysnonresZ\%$ \\
~~~MC statistics & $\sysMCKPPM$ & $\sysMCKSPP$ & $\sysMCKPPZ$ & $\sysMCKSPZ$ \\
\hline 
\end{mytable*}

\myfigure{0.5}{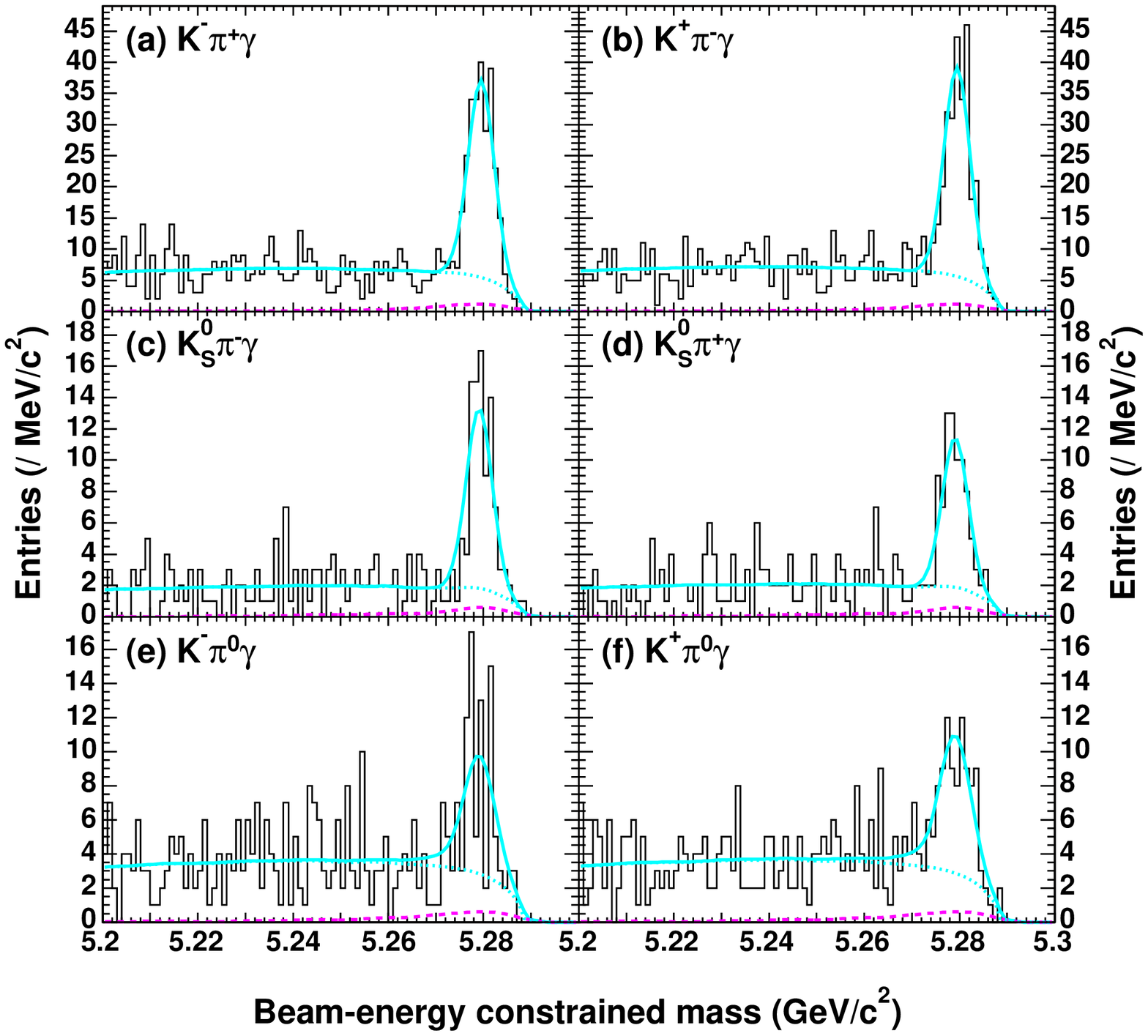}{%
  Fit results for the beam-energy constrained mass distributions for the
  search for partial rate asymmetry. The sum of the signal and the
  background components are shown in the solid curves, while the dotted
  and dashed curves represent the total backgrounds and the $B$ decay
  backgrounds, respectively. }



\begin{mytable}{lccc}{branching}{Results for the signal yields, efficiencies
and branching fractions ($\Br$).}
\hline 
{} & signal yield & efficiency & $\Br$ ($\times 10^{-5}$)\\
\hline 
$K^{*0}\gamma$ & $\NumKZG$ & $\EffKZG$ & $\BrBtoKZG$ \\
$K^{*+}\gamma$ & $\NumKPG$ & $\EffKPG$ & $\BrBtoKPG$ \\
\hline 
$(\KP\piM)\gamma$ & $\NumKPPMG$ & $\EffKPPMG$ & $\BrBtoKPPMG$ \\
$(\KS\piZ)\gamma$ & $\NumKSPZG$ & $\EffKSPZG$ & $\BrBtoKSPZG$ \\
$(\KS\piP)\gamma$ & $\NumKSPPG$ & $\EffKSPPG$ & $\BrBtoKSPPG$ \\
$(\KP\piZ)\gamma$ & $\NumKPPZG$ & $\EffKPPZG$ & $\BrBtoKPPZG$ \\
\hline 
\end{mytable}

\begin{mytable*}{lccc}{acp}{Results of the partial rate asymmetry search.}
\hline 
$K^*$ mode & $N(\Bbar\to\Kbar^*\gamma)$ & $N(B\to K^*\gamma)$ 
           & $\Acp$ \\
\hline 
$K^{*0}\to K^+\pi^-$ & $\NnegBtoKZG$ & $\NposBtoKZG$ & $\AcpBtoKZG$ \\
$K^{*+}\to \KS\pi^+$ & $\NnegBtoKSPPG$ & $\NposBtoKSPPG$ & $\AcpBtoKSPPG$ \\
$K^{*+}\to K^+\pi^0$ & $\NnegBtoKPPZG$ & $\NposBtoKPPZG$ & $\AcpBtoKPPZG$ \\
Combined ($K^{*+}$)   &                 &                 & $\AcpBtoKPG$ \\
\hline 
Combined (all)        &                 &                 & $\AcpBtoKG$ \\
\hline 
\end{mytable*}
\end{document}